\documentstyle[epsf,seceq]{ptptex}
\newcommand{\ymh}{Yang-Mills-Higgs Lagrangian}
\newcommand{\p}{\partial}
\newcommand{\nonum}{\nonumber\\}
\newcommand{\be}{\begin{equation}}
\newcommand{\ee}{\end{equation}}
\newcommand{\ba}{\begin{eqnarray}}
\newcommand{\ea}{\end{eqnarray}}
\newcommand{\bd}{\begin{displaymath}}
\newcommand{\ed}{\end{displaymath}}

\renewcommand{\a}{&&}
\newcommand{\ma}[1]{_{\mathop{}_{#1}}}

\makeatletter

          \@addtoreset{equation}{section}
       \makeatother
\title{%
Non-Commutative Differential Geometry\\
 on Discrete Space $M_4\times Z\ma{N}$ 
and Gauge Theory
}
\author{%
  Yoshitaka {\sc Okumura}\footnote{
e-mail address: okum@isc.chubu.ac.jp}
}
\inst{%
Department of Natural Science, 
Chubu University, Kasugai, 487, Japan
}
\abst{%
The algebra of non-commutative differential geometry (NCG) 
on the discrete space
$M_4\times Z\ma{N}$ previously proposed by the present author 
is improved to give the consistent explanation of the generalized
gauge field as the generalized connection on $M_4\times Z\ma{N}$.
The nilpotency of the generalized exterior derivative ${\bf d}$ 
is easily proved. The matrix formulation where the generalized gauge field
is denoted in matrix form is shown to have the same content with
the  ordinary formulation using {\bf d}, 
which helps us understand the implications
of the algebraic rules of NCG on $M_4\times Z\ma{N}$.
The Lagrangian of spontaneously broken gauge theory which has 
the extra restriction 
on the coupling constant of the Higgs potential is obtained by taking the 
inner product of the generalized field strength. The covariant derivative 
operating on the fermion field determines the parallel transformation
on $M_4\times Z\ma{N}$, which confirms that 
the Higgs field is  the connection on the discrete space. This implies
that the Higgs particle is a gauge particle on the same footing as
 the weak bosons.
Thus, it is expected that the mass relation 
$m\ma{H}=\frac{4}{\sqrt{3}}m\ma{W}\sin\theta\ma{W}$ proposed 
by the present author
holds without any correction in the same way as the mass relation 
$m\ma{W}=m\ma{Z}\cos\theta\ma{W}$.
The Higgs kinetic and potential terms are regarded as the curvatures 
on $M_4\times Z\ma{N}$.
}
\begin{document}
\maketitle
\section{Introduction}
In order to understand the essence of the Higgs mechanism 
we have developed the formulation based on the non-commutative differential 
geometry (NCG) on the discrete space $M_4\times Z_2$\cite{MO1}$^-$\cite{OSM} 
and in general $M_4\times Z\ma{N}$.\cite{O1}$^-$\cite{O10}
 Our formulation is a generalization
of ordinary differential geometry on a continuous manifold where the 
one-form connection on the principal bundle is defined, and then
 two-form curvature is deduced. If the structure group of the principal
bundle is non-abelian, the Lagrangian of 
the non-abelian gauge theory results from the
inner product of the curvature. We proceed in the same story as
that in the ordinary differential geometry though
our space is  discrete and a set of one form base consists of $dx^\mu$
and $\chi_k\;(k=1,2\sim N)$. This 
is the essential difference between 
our formulation and others such as Connes's\cite{Con}
 and Chamseddine, Felder and 
Fr\"olich's\cite{Cham} formulations. They use the Clifford algebra 
of the Dirac matrices $\gamma_\mu$ and $\gamma_5$ 
instead of $dx^\mu$ and $\chi_k$. 
We can define the exterior derivative satisfying
the nilpotency which is very important to obtain the gauge covariant 
formulation. This point is rather unclear in Refs.\citen{Con} and \citen{Cham}
though their works are pioneering. In any way, the Higgs field 
is regarded as a kind of gauge field on the discrete space in the same 
way as the ordinary gauge field, so that 
the raison d'${\hat e}$-tre
of Higgs field is apparent and any extra physical modes are not yielded. 
 \par
In this article, we revise the algebraic rule of $d_{\chi_k}$ operation
on $M_{nl}\chi_{\chi_l}$ which is an important factor
responsible for the symmetry breakdown. We can achieve easier proof
of the nilpotency of the generalized exterior derivative 
${\bf d}=d+d_\chi$ with $d_\chi=\sum_kd_{\chi_k}$ than that in Ref.\citen{O1}.
It should be noticed that in our formulation 
two-form basis $\chi_k\wedge\chi_l$ is independent of
$\chi_l\wedge\chi_k$ for $k\ne l$ contrary to the ordinary commutative algebra.
The name of our NCG results from this non-commutativity.
In addition, through the covariant derivative of fermion field,
this revision makes it possible to define the parallel
transformation of fermion field on the discrete space $Z\ma{N}$
in the same way as that on the ordinary continuous manifold. 
This definition confirms the Higgs field to be the connection on the discrete
space. 
Konisi and Saito\cite{KS} indicated that  the Higgs field
is a connection  governing the parallel transformation on the discrete space.
Our discussions are inspired by their indication.
\par
Chamseddine, Felder and Fr\"olich\cite{Cham} 
extended NCG formulation in the matrix form by use of 
the Dirac matrices $\gamma_\mu,\;\gamma_5$. 
We also show in this article that the same type matrix formulation is possible
by use of one-form base $dx^\mu$ and ${\chi_k}$ instead of the Dirac matrices.
${\bf d}$ operations on matrices are properly defined in conformity with
our formulation stated above. As a result, the nilpotency of ${\bf d}$
holds, which leads to the consistent gauge covariant formulation.
The corresponding operation in Ref.\citen{Cham} to ${\bf d}$ in ours
is the commutation relation of matrices and therefore the corresponding 
equation to the nilpotency of ${\bf d}$ doesn't hold, 
so that it is vague whether
their formulation is consistent or not. It seems that their formulation
is not applicable to $N\geq 4$ case because they have only one $\gamma_5$.
On the contrary, our method is applicable to any $N$ case 
because we have many $\chi_k\:(k=1,2\cdots N)$.
The shifting rule used in the algebraic calculation is very important
in our framework stated in \S 3. 
However, it is somewhat hard to understand 
the implication of that rule. The matrix formulation 
makes it very clear since  matrix manipulations bring 
the resultant equations after the shifting rule has already performed.
\par
This article consists of five sections and one appendix.
The next section presents the fundamental framework of NCG 
on $M_4\times Z\ma{N}$ with the emphasis on 
the nilpotency of ${\bf d}$. 
In third section, the matrix formulation of our NCG 
is developed, which leads to the same result as that in second section.
In fourth section, fermion sector of NCG is treated with the emphasis
on the parallel transformation of fermion field.
The last section is devoted to concluding remarks. 
The relationship between the NCG on $M_4\times Z_2$
stated in \S 2 and that of our previous framework with only one $\chi$ 
becomes clear in appendix A.
\section{  Non-commutative differential geometry 
on $M_4\times Z\ma{N}$}
The algebra of the previous formulation of NCG on 
the discrete space $M_4\times Z\ma{N}$\cite{O10}
is improved to yield more reasonable understanding  of
the generalized gauge field ${\cal A}(x,n)$
 in which  $x_\mu$  and $n$  are the arguments of fields 
 in  $M_4$ and  $Z\ma{N}$, respectively. 
\par
Let us start with the definition of the generalized gauge field 
${\cal A}(x,n)$
written in one-form on the discrete space $M_4\times Z\ma{N}$:
\be
      {\cal A}(x,n)
      =\sum_{i}a^\dagger_{i}(x,n){\bf d}
      a_i(x,n). \label{2.1}
\ee
$a_i(x,n)$ is the square-matrix-valued functions.
The subscript $i$ is  a variable of the extra
internal space which we can not now identify.  
Now, we simply regard $a_i(x,n)$  as the more fundamental
field from which to construct gauge field and the Higgs fields. They have only
mathematical meaning because $a_i(x,n)$ 
does not appear in final stage.
${\bf d}$ in Eq.(\ref{2.1}) is the generalized exterior 
derivative defined as follows.
\ba 
      \a {\bf d}=d + d_\chi=d+\sum_k d_{\chi_k}, \label{2.2.1}\\  
     \a da_i(x,n) = 
   \p_\mu a_i(x,n)dx^\mu, \label{2.2.2}\\
  \a d_{\chi_k} a_i(x,n)=[-a_i(x,n)M_{nk}\chi_k 
      + M_{nk}\chi_k a_i(x,n)], \label{2.2.3}
\ea
where $dx^\mu$ is a 
ordinary one-form basis, taken to be dimensionless, in Minkowski space 
$M_4$ and $\chi_k$ 
is the one-form basis, assumed to be also dimensionless, 
in the discrete space $Z\ma{N}$. 
We have introduced $x$-independent matrix $M_{nm}$ expressed in 
the rectangular matrix
whose hermitian conjugation is given by $M_{nk}^\dagger=M_{kn}$. 
If $a_i(x,n)$ and $a_i(x,k)$ are $L\times L$ and $K\times K$ square matrices,
 respectively, $M_{nk}$ is $L\times K$ type matrix. 
 $M_{nn}=0$ is assumed but it should be noted that this
equation must not be used until the calculation of $d_{\chi_k}$ operation
 is finished because it contradicts with the shifting rule as stated later.
The matrix $M_{nk}$ turns out to determine the scale and pattern of 
the spontaneous breakdown of the gauge symmetry. Thus, the symmetry
breaking mechanism is encoded in the $d_\chi$ operation.
Before finding the explicit forms of gauge and
the Higgs fields and generalized field strength according to Eqs.(\ref{2.1}) 
and (\ref{2.2.1})$\sim$(\ref{2.2.3}), 
we determine the several important algebraic rules
 in non-commutative geometry.
We first investigate the condition that the Leibniz rule 
with respect to ${ d_{\chi_k}}$ consistently holds. 
The Leibniz rule is written as (the subscript $i$ of
$a_i(x,n)$ is abbreviated for a while)
\be
  { d_{\chi_k}}(a(x,n)b(x,n))
  =({ d_{\chi_k}}a(x,n))b(x,n)+a(x,n)({d_{\chi_k}}b(x,n)), 
  \label{2.3}
\ee
where the $b(x,n)$ is also the same type 
matrix  as $a(x,n)$.
We have to assume the rather unusual rule for $\chi_k$ that
\be
            \chi_k b(x,n) = b(x,k)\chi_k,  \label{2.4.1}
\ee
which makes the product of matrix valued functions consistently calculable.
Eq.(\ref{2.4.1}) is only an example 
to show how to change the discrete variable 
 when $\chi_k$ is shifted from left to right in an equation. 
For example,
\be
           \chi_k M_{nl}=M_{kl}\chi_k.  \label{2.4.2}
\ee
Such a change is necessary to assure the consistent product of matrices.
We call this rule the shifting rule which reflects the noncommutative 
nature of the differential calculus just now treated.
This rule is reasonably understood and its justification is confirmed 
in the matrix formulation of NCG 
described in the next section.
According to this shifting rule, the third equation in Eq.(\ref{2.2.3}) 
is rewritten as
\be
   d_{\chi_k} a(x,n) = [-a(x,n)M_{nk} + M_{nk}a(x,k)]\chi_k\,. \label{2.5}
\ee
Eq.(\ref{2.5}) and the shifting rule 
enable us to prove the Leibniz rule for $d_\chi$ in Eq.(\ref{2.3}).
\par
We next investigate the nilpotency of ${\bf d}$ which is very important
to obtain the generalized field strength. 
From the definition of ${\bf d}$ in Eq.(\ref{2.2.1}),
\be
 {\bf d}^{\,2} a(x,n)
 =(d^{\,2}+ d\cdot d_\chi +d_\chi\cdot d +d_\chi^{\,2})a(x,n). \label{2.6}
\ee
$d^{\,2}=0$ naturally holds in ordinary differential forms.
$dx^\mu\wedge \chi_k=-\chi_k\wedge dx^\mu$
is reasonably assumed  and it yields
\be
      (d\cdot d_\chi +d_\chi\cdot d)a(x,n)=0. 
\label{2.7}
\ee
However, $d_\chi^{\,2}a(x,n)=0$ is not so easily proved
that we need to add a following algebraic rule\footnote{This rule is revised
from that in Ref.\citen{O1} which was written as
\ba
  d_{\chi_k} (M_{nl}\chi_l) \a= M_{nl}M_{lk}\chi_{l}\wedge \chi_k. \nonumber
\ea
This revision yields  simpler proof of the nilpotency of ${\bf d}$, 
which is only one difference between this section and that in Ref.\citen{O1}.}
 for $d_\chi$ operation on $M_{nk}$:
\be
         d_{\chi_l}M_{nk}\chi_k=M_{nl}\chi_l\wedge M_{nk}\chi_k=
         M_{nl}M_{lk}\chi_l\wedge\chi_k.
         \label{2.8}
\ee
In addition, the following rule is taken into account; whenever the 
$d_{\chi_k}$ operation jumps over $M_{nl}\chi_l$, minus sign attached,
for example
\be
    d_{\chi_k}\left(M_{nl}\chi_la(x,n)\right)=
    \left(d_{\chi_k}M_{nl}\chi_l\right)a(x,n)
    - M_{nl}\chi_l\wedge \left(d_{\chi_k}a(x,n)\right).
\label{2.9}
\ee
In this equation, we should notice that
$\chi_l\wedge\chi_k$ is independent of $\chi_k\wedge\chi_l$ for $k\ne l$
 due to the noncommutative property of geometry on the discrete
space. 
With these considerations, we can easily calculate: 
\ba
    d_{\chi_l}\left(d_{\chi_k}a(x,n)\right)
  = \a d_{\chi_l}\left(-a(x,n)M_{nk}\chi_k+M_{nk}\chi_k a(x,n)\right)\nonum
  =\a -\left(d_{\chi_l}a(x,n)\right)\wedge M_{nk}\chi_k
    -a(x,n)\left(d_{\chi_l}M_{nk}\chi_k\right)\nonum
  \a \hskip 1.5cm +\left(d_{\chi_l}M_{nk}\chi_k\right)a(x,n)
    -M_{nk}\chi_k \wedge \left( d_{\chi_l}a(x,n)\right)\nonum
 =\a\left(-M_{nl}a(x,l)M_{lk}+M_{nl}M_{lk}a(x,k)\right)\chi_l\wedge\chi_k\nonum    \a\hskip 0.5cm
 -\left(-M_{nk}a(x,k)M_{kl}+M_{nk}M_{kl}a(x,l)\right)\chi_k\wedge\chi_l,
  \label{2.10}
\ea
which yields 
\be
 \left(d_{\chi_l}d_{\chi_k}+d_{\chi_k}d_{\chi_l}\right)a(x,n)=0.
 \label{2.11}
\ee
From Eq(\ref{2.11}) we obtain the nilpotency of $d_\chi^{\,2}a(x,n)=0$.
Then, according to Eqs.(\ref{2.6}) and (\ref{2.7}), the nilpotency of
${\bf d}^{\,2}a(x,n)=0$ is followed.
In the similar way, we can prove 
\be
 \left(d_{\chi_l}d_{\chi_k}+d_{\chi_k}d_{\chi_l}\right)M_{nm}\chi_m=0,
 \label{2.12}
\ee
which together with Eq.(\ref{2.9}) 
helps us prove, for example, $ d_\chi^{\,2}\{ M_{nm}\chi_m a(x,m)\}=0$.
However, we don't need such a general case 
but only quote ${\bf d}^{\,2}a(x,n)=0$ in this article.\par
Inserting Eqs.(\ref{2.2.1})$\sim$(\ref{2.2.3}) into Eq.(\ref{2.1})
and using Eq.(\ref{2.3}),
${\cal A}(x,n)$ is rewritten as
\be
 {\cal A}(x,n)={A}_\mu(x,n)dx^\mu
               +\sum_k{\mit\Phi}_{nk}(x)\chi_k, \label{2.13}
\ee
where
\ba
&& A_\mu(x,n) = \sum_{i}a_{i}^\dagger(x,n)\p_\mu a_{i}(x,n), \label{2.14.1}\\
&&     {\mit\Phi}_{nk}(x) = \sum_{i}a_{i}^\dagger(x,n)
         \,[-a_i(x,n)M_{nk} 
            + M_{nk}a_i(x,k)], \label{2.14.2}
\ea
$A_\mu(x,n)$ and ${\mit\Phi}_{nk}(x)$, 
 are identified with
the gauge field in the flavor symmetry and the Higgs field,  respectively. 
In order to identify  $A_\mu(x,n)$ 
with true gauge fields, the following condition has to be imposed:
\ba
&&    \sum_{i}a_{i}^\dagger(x,n)a_{i}(x,n)= 1. 
  \label{2.15}
\ea
\par
Before constructing the gauge covariant field strength, 
we address the gauge transformation 
of $a_i(x,n)$  which is defined as 
\ba
&&      a^{g}_{i}(x,n) = a_{i}(x,n)g(x,n), 
\label{2.16}
\ea
where
$g(x,n)$ is the gauge function with respect to the corresponding
flavor unitary group specified by $n$. Let us define the $d_\chi$
operation on $g(x,n)$ by
\ba 
       {\bf d}g(x,n)\a=(d+\sum_kd_{\chi_k}) g(x,n) \nonum
             \a = \p_\mu g(x,n)dx^\mu
             +\sum_k[-g(x,n)M_{nk} + M_{nk}g(x,k)]\chi_k.
          \label{2.17}
\ea
Then, we can find from Eq.(\ref{2.1}) 
the gauge transformation of ${\cal A}(x,n)$ to be
\be
{\cal A}^g(x,n)=g^{-1}(x,n) 
{\cal A}(x,n)g(x,n)  +g^{-1}(x,n){\bf d}g(x,n), \label{2.18}
\ee
which using  (\ref{2.14.1}) and (\ref{2.14.2}) leads to 
 the gauge transformations of gauge, and Higgs fields as
\ba
\a      \hskip-1cm A_\mu^g(x,n)=g^{-1}(x,n)
   A_\mu(x,n)g(x,n)+ g^{-1}(x,n)\p_\mu g(x,n),   \label{2.19}\\
\a  \hskip -1cm {\mit\Phi}^g_{nk}(x)=g^{-1}(x,n){\mit\Phi}_{nk}(x)
g(x,k)+  g^{-1}(x,n)[- g(x,n)M_{nk}+M_{nk}g(x,k)], \label{2.20}
\ea
Equation(\ref{2.20}) is very similar to Eq.(\ref{2.19}) 
if the inhomogeneous term is written as
\be
    g^{-1}(x,n)[-g(x,n)M_{nk}+M_{nk}g(x,k)]=g^{-1}(x,n)\p_{nk}g(x,n).
    \label{2.21}
\ee
In this context, $\p_{nk}$ corresponds to $\p_\mu$. 
Thus, $\p_{nk}$ seems to be the difference operator on the discrete space.
$M_{nk}$ in Eq.(\ref{2.20}) is inserted to insure the consistent calculation
of matrix products.
The resemblance between Eqs.(\ref{2.19}) and (\ref{2.20}) 
 strongly indicates that the Higgs field is a kind of gauge field
on the discrete space $M_4\times Z\ma{N}$. 
 Equation(\ref{2.20}) is rewritten as
\be
       {\mit\Phi}^g_{nk}(x)+M_{nk}=g^{-1}(x,n)
       ({\mit\Phi}_{nk}(x)+M_{nk})g(x,k),
                              \label{2.22}
\ee
which makes it obvious that 
\be
H_{nk}(x)={\mit\Phi}_{nk}(x)+M_{nk} \label{2.23}
\ee
is un-shifted Higgs field whereas ${\mit\Phi}_{nk}(x)$ 
denotes shifted one with the vanishing vacuum expectation value.
\par
With these considerations we can construct the gauge covariant field
strength:
\be
  {\cal F}(x,n)= {\bf d}{\cal A}(x,n) +{\cal A}(x,n)\wedge{\cal A}(x,n)
\label{2.24}
\ee
According to the nilpotency of ${\bf d}$, we can write
$$
{\bf d}{\cal A}(x,n)
=\sum_i{\bf d}a_i^\dagger(x,n)\wedge {\bf d}a_i^\dagger(x,n)
$$
which along with Eqs.(\ref{2.2.2}), (\ref{2.2.3}) 
and (\ref{2.13}) serves us to introduce
the explicit expression of ${\cal F}(x,n)$ as
\ba
 {\cal F}(x,n) &=& { 1 \over 2}F_{\mu\nu}(x,n)dx^\mu \wedge dx^\nu + 
               \sum_{k\ne n}D_\mu\Phi_{nk}(x)dx^\mu \wedge \chi_k \nonum
             & & + \sum_{k\ne n}V_{nk}(x)\chi_k \wedge \chi_n \nonum
     && + \sum_{k\ne n}\sum_{\{l\ne k,n\}}V_{nkl}(x)\chi_k \wedge \chi_l, 
                \label{2.25}
\ea
where
\ba
 \a \hskip-1cm
 F_{\mu\nu}(x,n)=\partial_\mu A_\nu (x,n) - \partial_\nu A_\mu (x,n) 
               + [A_\mu(x,n), A_\mu(x,n)], \label{2.26}\\
  \a \hskip -1cm D_\mu\Phi_{nl}(x)=\partial_\mu \Phi_{nl}(x) - 
  (\Phi_{nk}(x)+M_{nk})A_\mu(x,k)\nonum
        && \hskip 5.5cm + A_\mu(x,n)(M_{nk} + \Phi_{nk}(x)),\label{2.27}\\
 \a \hskip-1cm V_{nk}(x)= (\Phi_{nk}(x) + M_{nk})(\Phi_{kn}(x) + M_{kn}) -
             Y_{nk}(x),\hskip 0.5cm {\rm for}\,\,k\ne n,\label{2.28}\\
 \a \hskip-1cm V_{nkl}(x)= (\Phi_{nk}(x) + M_{nk})(\Phi_{kl}(x) + M_{kl}) -
             Y_{nkl}(x), \hskip 0.5cm{\rm for}\,\,k\ne n, \,\, l\ne n.
     \label{2.29}
\ea
\par
$Y_{nk}(x)$ and $Y_{nkl}(x)$ are the auxiliary fields written as
\ba
  \a Y_{nk}(x)=\sum_i a^\dagger_i(x,n)M_{nk}M_{kn}a_i(x,n), \label{2.29a1}\\
      \a Y_{nkl}(x)=\sum_i a^\dagger_i(x,n)M_{nk}M_{kl}a_i(x,l). \label{2.29a2}
\ea

In order to obtain the gauge invariant \ymh, we address the
gauge transformation of ${\cal F}(x,n)$.
From Eq.(\ref{2.18})  we can easily find the gauge 
transformation of ${\cal F}(x,n)$ as
\be
         {\cal F}^g(x,n)=g^{-1}(x,n){\cal F}(x,n)g(x,n).  \label{2.30}
\ee
The metric structure of one-forms are defined as
\be
<dx^\mu, dx^\nu>=g^{\mu\nu},\hskip 1cm 
<\chi_n, dx^\mu>=0, \hskip 1cm
<\chi_n, \chi_k>=-\alpha^2\delta_{nk},
\label{2.31}
\ee
which determines the inner products of two-forms
such as
\ba
\a <dx^\mu \wedge dx^\nu,
dx^\rho \wedge dx^\sigma>=g^{\mu\rho}g^{\nu\sigma}-
g^{\mu\sigma}g^{\nu\rho}, \label{2.32}\\
\a <dx^\mu \wedge \chi_n,
dx^\nu \wedge \chi_k>=-\alpha^2g^{\mu\nu}\delta_{nk},\label{2.33}\\
\a <\chi_n \wedge \chi_k,\chi_m \wedge \chi_l>
      =\alpha^4 \delta_{nm}\delta_{kl}. \label{2.34}
\ea
while other inner products among the basis two-forms vanish. 
It should be noted that $\,\chi_m \wedge \chi_l\,$ is a two-form basis
independent of
 $\,\chi_l \wedge \chi_m\,$ for $m \ne l\,$, which reflects 
 the non-commutative property of geometry on the discrete space. 
Thus,  Eq.(\ref{2.34}) is followed. In this paper, we assume the coefficient
of $\delta_{nm}\delta_{kl}$ to be
$\alpha^4$ on the right-hand side in Eq.(\ref{2.34}). In general, it seems that
the coefficient can not be determined only by NCG. 
\par
According to these metric structures and Eq.(\ref{2.25}) we can obtain 
the formula for gauge-invariant \ymh
\ba
{\cal L}_{{\mathop{}_{\rm YMH}}}(x)
=\a-\sum_{n=1}^{\mathop{}_{N}}{1 \over g_{n}^2}
< {\cal F}(x,n),  {\cal F}(x,n)>\nonum
=\a-{\rm Tr}\sum_{n=1}^{\mathop{}_{N}}{1\over 2g^2_n}
F_{\mu\nu}^{\dag}(x,n)F^{\mu\nu}(x,n)\nonum
\a+{\rm Tr}\sum_{n=1}^{\mathop{}_{N}}\sum_{k\ne n}{\alpha^2\over g_{n}^2}
    (D_\mu\Phi_{nk}(x))^{\dag}D^\mu\Phi_{nk}(x)\nonum
-{\rm Tr}\sum_{n=1}^{\mathop{}_{N}}\a
{\alpha^4\over g_{n}^2}\sum_{k\ne n}V_{nk}^{\dag}(x)
V_{nk}(x)
-{\rm Tr}\sum_{n=1}^{\mathop{}_{N}}{\alpha^4\over g_{n}^2}
         \sum_{k\ne n}\sum_{l\ne n,k}V_{nkl}^{\dag}(x)V_{nkl}(x),
\label{2.35}
\ea
where $g_n$ is the gauge coupling constant
and
Tr denotes the trace over internal symmetry matrices. 
The third term is the potential term of Higgs particle and the fourth term
is the interaction term between Higgs particles. 
Thus, the fourth term does not appear when $N=2$.
\par
\section{ Matrix formulation of NCG}
In this section we explain the formulation on the discrete space
$M_4\times Z_3 \,(N=3)$ since it is very easy to generalize it 
to that on $M_4\times Z\ma{N}$. In addition,  $x_\mu$ as the argument 
in fields is abbreviated for simplicity. The fundamental field $a_i$
is introduced as
\be
     a_i={\rm diag}(a_i(1), a_i(2), a_i(3)), \label{3.1}
\ee
where $a_i(n)\,(n=1,2,3)$ is the same fundamental field as in Eq.(\ref{2.1}).
The generalized gauge field ${\cal A}$ is defined as
\be
     {\cal A}=\sum_ia^\dagger_i{\bf d}a_i, \label{3.2}
\ee
with the generalized exterior derivative ${\bf d}$ whose operation on
$a_i$ is described as
\be
        {\bf d}a_i= (d+d_\chi) a_i=da_i+d_\chi a_i.  \label{3.3}
\ee
$da_i$ in Eq.(\ref{3.3}) 
is explicitly written in matrix form as
\be
     da_i=\left(\matrix{\p_\mu a_i(1)dx^\mu & 0 & 0 \cr
                          0   & \p_\mu a_i(2)dx^\mu & 0 \cr
                          0  &  0 & \p_\mu a_i(3)dx^\mu \cr}\right)
                           \label{3.4}
\ee
and  $d_\chi a_i$ is defined as
\be
      d_\chi a_i=-a_i M +M a_i, \label{3.5}
\ee
where  $M$ is given as the matrix with one-form base $\chi_k\,(k=1,2,3)$:
\be
     M=\left(\matrix{0 & M_{12}\chi_2 & M_{13}\chi_3 \cr
                 M_{21}\chi_1   & 0 & M_{23}\chi_3 \cr
                 M_{31}\chi_1  &  M_{32}\chi_2 & 0 \cr}\right).
                           \label{3.6}
\ee
Then, we can find the matrix element of $d_\chi a_i$ as
\ba
     \a   (d_\chi a_i)_{nk}=-a_i(n)M_{nk}\chi_k+M_{nk}\chi_k a_i(k), \quad
       {\rm for} \quad n\ne k, \nonum
    \a    (d_\chi a_i)_{nn}=0. \label{3.7}
\ea
With these equations, we obtain the explicit form of ${\cal A}$ as
\be
     {\cal A}=\left(\matrix{
     A_\mu(1)dx^\mu & {\mit \Phi}_{12}\chi_2 & {\mit\Phi}_{13}\chi_3 \cr
     {\mit \Phi}_{21}\chi_1   & A_\mu(2)dx^\mu  & {\mit\Phi}_{23}\chi_3 \cr
 {\mit\Phi}_{31}\chi_1  &  {\mit\Phi}_{32}\chi_2 & A_\mu(3)dx^\mu  \cr}\right),
                           \label{3.8}
\ee
where the gauge field $A_\mu(k)\,(k=1,2,3)$ and the Higgs field
${\mit\Phi}_{nk}\, (n,k=1,2,3)$ are denoted in the same form as 
in Eqs.(\ref{2.14.1}) and (\ref{2.14.2}), respectively.
It should be noted that in introducing Eq.(\ref{3.8}) the shifting rule
in Eqs.(\ref{2.4.1}) or (\ref{2.4.2}) is not necessary because
in the matrix formulation of NCG, the expressions such as
\be
      M_{nk}\chi_k a_i(k)=M_{nk}a_i(k)\chi_k,\hskip 1.0cm 
      M_{nk}\chi_k M_{kl}=M_{nk}M_{kl}\chi_k   \label{3.9}
\ee
appear. Eq.(\ref{3.9}) has the forms 
after the shifting rule has been already  applied. 
 This fact makes the meaning of the shifting rule 
 expressed in Eqs.(\ref{2.4.1}) and (\ref{2.4.2}) very clear
 and confirms the justification of that rule. 
\par
The gauge transformation is given as
\be
     a_i^g=a_ig,   \label{3.11}
\ee
where the gauge transformation function $g$ is
expressed in matrix form as
\be
      g={\rm diag}(g(1),g(2),g(3)) \label{3.12}
\ee
and the $d_\chi$ operation on it is
\be
       d_\chi g =-gM+Mg=-[g, M]. \label{3.13}
\ee
From these equations, the gauge transformation of ${\cal A}$ is given as
\be
     {\cal A}^g =g^{-1}{\cal A}g + g^{-1}{\bf d}g \label{3.14}
\ee
which yields the same gauge transformations of $A_\mu(n)\,(n=1,2,3)$ and
${\mit\Phi}_{nk}\,(n,k=1,2,3)$ as those in Eqs.(\ref{2.19}) and (\ref{2.20}),
respectively.
\par
The generalized field strength is in this matrix formulation written as
\be
     {\cal F}={\bf d}{\cal A}+{\cal A}\wedge {\cal A}. \label{3.15}
\ee
In obtaining the explicit form of ${\cal F}$ we must again address
the nilpotency of ${\bf d}$. $(dd_\chi+d_\chi d)a_i=0$ easily results
under the reasonable condition that $dx^\mu\wedge \chi_k=-\chi_k\wedge dx^\mu$.
With the definition 
\be
      d_\chi M=M\wedge M   \label{3.16}
\ee
$d_\chi^{\,2}a_i$ is calculated as follows:
\ba
      d_\chi(d_\chi a_i)=\a d_\chi(-a_iM+Ma_i)\nonum
              =\a-(d_\chi a_i)\wedge M-a_i(d_\chi M)+(d_\chi M)a_i
              -M\wedge(d_\chi a_i)\nonum
   =\a -(-a_iM+Ma_i)\wedge M-a_i(M\wedge M) +(M\wedge M)a_i
     -M\wedge(-a_i M+Ma_i)\nonum
   =\a 0, \label{3.17}
\ea
where 
the rule is again adopted that whenever the 
$d_{\chi}$ operation jumps over $M$, minus sign attached.
We call this rule jumping rule, hereafter.
Then, we can rewrite Eq.(\ref{3.15}) as
\be
     {\cal F}=\sum_i{\bf d}a^\dagger_i\wedge{\bf d}a_i
        +{\cal A}\wedge {\cal A}. \label{3.18}
\ee
With the help of Eqs.(\ref{3.7}) and (\ref{3.8}) we can find  
the matrix elements of ${\cal F}$ as
\ba
      {\cal F}_{nn}=\a\frac12 F_{\mu\nu}(n)dx^\mu\wedge dx^\nu
    +\sum_{k\ne n}(H_{nk}H_{kn}-Y_{nk})\chi_k\wedge\chi_n
          \label{3.19}
\ea
and
\ba
      {\cal F}_{nk}=\a (\p_\mu H_{nk}+A_\mu(n)H_{nk}-H_{nk}A_\mu(k)
                )dx^\mu\wedge \chi_k\nonum
    \a\hskip 2cm+\sum_{l\ne n,k}(H_{nl}H_{lk}-Y_{nlk})\chi_l\wedge\chi_k,
         \quad {\rm for}\quad n\ne k.
          \label{3.20}
\ea
$H_{nk}$, $Y_{nk}$ and $Y_{nlk}$ are written in Eqs.(\ref{2.23}),
(\ref{2.29a1}) and (\ref{2.29a2}), respectively.
\par
With the same metric structures in Eqs.(\ref{2.32})$\sim$(\ref{2.34}), 
we can obtain the \ymh $\;$from the following equation;
\be
     L_{_{\rm YMH}}=-{\rm Tr}<{\cal F}, C {\cal F}>, \label{3.21}
\ee
where $C$ is a matrix relating to the coupling constant 
which is given as
\be
       C={\rm diag}(\frac1{g_{1}^2},\frac1{g_{2}^2},\frac1{g_{3}^2}).
       \label{3.22}
\ee
It should be noted that ${\rm Tr}$ in Eq.(\ref{3.21}) 
is the trace not only over the $3\times 3$ matrix such as in Eq.(\ref{3.8})
but also over the internal symmetry matrices.
It is easily understand that Eq.(\ref{3.21}) yields the same expression
as that in Eq.(\ref{2.35}).
\section{ Fermion sector and parallel transformation}
We have already performed in the previous paper \cite{O1}
to reconstruct the Dirac Lagrangian in NCG by use of the differential
forms in the same way as the Yang-Mills-Higgs Lagrangian in Eq.(\ref{2.35}).
However, 
it was not done to discuss the parallel transformation of fermion 
field because the algebra of NCG  on $M_4\times Z\ma{N}$ was
not fully satisfactory. In this section, we review how to construct
the Dirac Lagrangian and discuss the parallel transformation 
of fermion field based on the revised algebra stated in \S 2.
In addition, the matrix formulation of the fermion sector is also
presented.\par
The covariant derivative acting on the spinor $\psi(x,n)$ is defined by
\be
{\cal D}\psi(x,n)=({\bf d}+ A^f(x,n))\psi(x,n), \label{4.1}
\ee
where $A^f(x,n)$
is the differential representation with respect to the fermions
such that
\be
{\cal A}^f(x,n)
=A_\mu^f(x,n)dx^\mu+\sum_m {\mit\Phi}^{f}_{nm}(x)\chi_m. \label{4.2}
\ee
It should be noticed that $ A_\mu^f(x,n)$ 
is the differential representation 
for $\psi(x,n)$ and
$\Phi_{nm}^{f}(x)$ also has  the expression corresponding  
to  $\psi(x,n)$. 
However, in almost all model building we can construct ${\cal A}(x,n)$ 
in \S 3 so as to coincide with ${\cal A}_\mu^f(x,n)$ in Eq.(\ref{4.2}).
Thus, the superscript $f$ in Eq.(\ref{4.2}) is removed, hereafter.
We further define $d_\chi$ operation on $\psi(x,n)$ as
\ba
d_\chi \psi(x,n)\a=\sum_k d_{\chi_k} \psi(x,n)\nonum
     \a=\sum_k M_{nk}\chi_k\psi(x,n)
     =\sum_k M_{nk}\psi(x,k)\chi_k ,
 \label{4.3}
\ea
which helps us rewrite Eq.(\ref{4.1}) as
\be
{\cal D}\psi(x,n)
= \left(\partial_\mu + A_\mu(x,n)dx^\mu+\sum_k H_{nk}(x)\chi_k\right)\psi(x,n),
\label{4.4}
\ee
with $H_{nk}(x)$ in Eq.(\ref{2.23}).\par
\par
Since the covariant derivative of fermion field has been defined 
in Eq.(\ref{4.1}), we can
address the parallel transformation of fermion field on the discrete 
space $M_4\times Z\ma{N}$.
If Eq.(\ref{4.3}) is denoted as
\be
d_\chi \psi(x,n)=\sum_k\p_{nk}\chi_k\psi(x,n), \label{4.7}
\ee
the covariant derivative ${\cal D}\psi(x,n)$ is rewritten as
\ba
  \hskip -1cm {\cal D}\psi(x,n)=\a\left\{(\p_\mu+A_\mu(x,n))dx^\mu\psi(x,n)
         +\sum_k(\p_{nk}+{\mit\Phi}_{nk}(x))\chi_k\psi(x,n)
         \right\}.   \label{4.8}
\ea
Eq.(\ref{4.8}) implies that ${\mit\Phi}_{nk}(x)\chi_k\psi(x,n)$ 
expresses the variation accompanying the parallel transformation
from $k$-th to $n$-th points on the discrete space just as\\
$A_\mu(x,n)dx^\mu\psi(x,n)$ is the variation of parallel
transformation on the Minkowski space $M_4$.
This makes it determinant that the shifted Higgs field $\mit \Phi_{nk}(x)$ 
is the gauge field on  the discrete space.\par
The nilpotency of $d_\chi$ in this case is also important
to obtain the consistent explanations of covariant derivative and 
parallel transformation. From Eq.(\ref{4.3}), we can easily calculate
\ba
  d_{\chi_l}(d_{\chi_k}\psi(x,n))=\a d_{\chi_l}(M_{nk}\chi_k\psi(x,n))\nonum
           =\a (d_{\chi_l}M_{nk}\chi_k)\psi(x,n)
           -M_{nk}\chi_k\wedge d_{\chi_l}\psi(x,n)\nonum
        =\a M_{nl}\chi_l\wedge M_{nk}\chi_k\psi(x,n)-
        M_{nk}\chi_k\wedge M_{nl}\chi_l\psi(x,n) \label{4.9}
\ea
which leads to
\be
    (d_{\chi_l}d_{\chi_k}+d_{\chi_k}d_{\chi_l})\psi(x,n)=0. \label{4.10}
\ee
Equation(\ref{4.10}) implies that if the Higgs field ${\mit\Phi}_{nk}(x)$
as the gauge field on the discrete space
becomes vanishing, the curvature on the discrete space also vanishes.
In Eq.(\ref{4.9}), it should be again noted that two-form basis
$\chi_l\wedge\chi_k$ is independent of $\chi_k\wedge\chi_l$ for $k\ne l$
because of the noncommutativity of our algebra.
From Eq.(\ref{4.10}) the nilpotency of ${\bf d}$ is evident.
Thus, we can obtain the generalized field strength as the 
curvature on the discrete space $M_4\times Z\ma{N}$ as follows:
\ba
{\cal F}(x,n)=\a({\bf d}+{\cal A}(x,n))\wedge ({\bf d}+{\cal A}(x,n))\nonum
     =\a {\bf d}{\cal A}(x,n)+{\cal A}(x,n)\wedge {\cal A}(x,n). \label{4.11}
\ea
Thus, we can regard the Higgs kinetic term in Eq.(\ref{2.27}),
 the Higgs potential term in Eq.(\ref{2.28}) and the Higgs interacting 
term in Eq.(\ref{2.29}) as the curvatures 
accompanying the parallel transformation on $M_4\times Z\ma{N}$.
\par
In the follows, we investigate the gauge transformation property 
of ${\cal D}\psi(x,n)$.
Let the gauge transformation of $\psi(x,n)$ to be 
\be
   \psi^g(x,n) = g^{-1}(x,n) \psi(x,n), \label{4.5}
\ee
with the same gauge transformation function in Eq.(\ref{2.16}).
Putting an attention on the equation that 
${\bf d}g(x,n)^{-1}=-g^{-1}(x,n)({\bf d}g(x,n)) g^{-1}(x,n)$,
we easily obtain 
the gauge covariance of ${\cal D}\psi(x,n)$.
\be
     {\cal D}\psi^g(x,n)=g^{-1}(x,n){\cal D}\psi(x,n). \label{4.6}
\ee
\par
In order to get the Dirac Lagrangian by use of the inner products 
of differential forms such as Eq.(\ref{2.35})
we introduce
the associated spinor one-form
by
\be
{\tilde {\cal D}}\psi(x,n)= \gamma_\mu \psi(x,n)dx^\mu
              -i{c_{ \mathop{}_{Y}}}\psi(x,n)\sum_m\chi_m,
\label{4.12}
\ee
where ${c\ma{Y}}$ is a real,
dimensionless constant and invariant
against the gauge transformation. 
It is evident that ${\tilde {\cal D}}\psi(x,n)$ is gauge covariant
\be
      {\tilde {\cal D}}\psi^{g}(x,n) 
      =g^{-1}(x,n){\tilde {\cal D}}\psi(x,n).        \label{4.13}
\ee
The Dirac Lagrangian invariant under the Lorentz and gauge transformations
 is obtained by taking the inner product
\ba
{\cal L}_{ \mathop{}_{D}}(x,n)=\a 
i{\rm Tr}\,<{\tilde {\cal D}}\psi(x,n),{\cal D}\psi(x,n)>\nonum
=\a i\,{\rm Tr}\,
[\,{\bar\psi}(x,n)\gamma^\mu(\partial_\mu+A_\mu(x,n))\psi(x,n)
+i{g_{ \mathop{}_{Y}}}{\bar\psi}(x,n)\sum_m H_{nm}(x)\psi(x,m)\,],\nonum
\a\label{4.14}
\ea
where ${-c_{ \mathop{}_{Y}}}\alpha^2$ 
is replaced simply by ${g_{ \mathop{}_{Y}}}$, and 
we  have used the definitions of the inner products for spinor one-forms
due to Eq.(\ref{2.31}),
\ba
\a <A(x,n)dx^\mu, B(x,n)dx^\nu>={\rm Tr}\,\bar{A}(x,n)B(x,n)g^{\mu\nu},
\label{4.15}\\
\a <A(x,n)\chi_k, B(x,n)\chi_l>
                  =-\alpha^2\delta_{kl}{\rm Tr}\,\bar{A}(x,n)B(x,n),
                  \label{4.16}
\ea
with vanishing other inner products.
The total Dirac Lagrangian is the sum over $n=1,2\cdots N$.
\be
{\cal L}_{\mathop{}_{D}}(x)
      =\sum_{n=1}^{\mathop{}_{N}}{\cal L}_{\mathop{}_{D}}(x,n), 
\label{4.17}
\ee
which  evidently satisfies the Hermiticity condition.\par
We can develop the similar discussions also in matrix formulation of NCG
stated in \S 3. Fermion field is expressed in the same notation in \S 3 as 
\be
      \psi=\left(\matrix{ \psi(1)\cr
                           \psi(2)\cr
                           \psi(3)\cr}\right).   \label{4.18}
\ee
The covariant derivative ${\cal D}\psi$ is described to be 
\be
          {\cal D}\psi=({\bf d}+{\cal A})\psi. \label{4.19}
\ee
Since $d_\chi$ operation on $\psi$ is
\be
                  d_\chi \psi=M\psi,         \label{4.20}
\ee
the nilpotency of $d_\chi$ is easily proved by
\be
       d_\chi(d_\chi \psi)=d_\chi(M\psi)=(d_\chi M)\psi-M(d_\chi \psi)
          =(MM)\psi-M(M\psi)=0,      \label{4.21}
\ee
where the jumping rule is applied. Thus, the nilpotency of ${\bf d}$
is followed. From these considerations, the generalized field strength
${\cal F}$ is written as
\ba
          {\cal F}=\a({\bf d}+{\cal A})\wedge({\bf d}+{\cal A})\nonum
                  =\a {\bf d}{\cal A}+{\cal A}\wedge{\cal A} \label{4.22}
\ea
which amounts to Eq.(\ref{3.18}).\par
Though the associated spinor one-form in Eq.(\ref{4.12}) has rather unclear
  expression in matrix notation, we define it so as to 
  have the elements expressed in Eq.(\ref{4.12}).
Then, we can obtain the Dirac Lagrangian as follows:
\be
{\cal L}_{ \mathop{}_{D}}= 
i{\rm Tr}\,<{\tilde {\cal D}}\psi,{\cal D}\psi>,
\label{4.23}
\ee
which is equal to that in Eq.(\ref{4.17}). Therefore, 
 we obtain 
 the same results also in the fermion sector as those
in the non-matrix formulation.
\par
\section{ Concluding remarks}
The algebra of NCG on $M_4\times Z\ma{N}$ previously proposed by the 
present author\cite{O1} was improved by revising the ${\chi_k}$
operation on $M_{nl}\chi_l$. This revision makes it possible that
the nilpotency of the exterior derivative ${\bf d}$ is easily proved, and
then the parallel transformation of fermion field is defined 
through the covariant derivative of fermion field.
The discussions of the parallel transformation make it 
decisive that the Higgs field is a connection on the discrete space 
in the same way as the ordinary gauge field on $M_4$ and 
the Higgs kinetic, potential and interacting terms in Lagrangian
are the curvatures accompanying the parallel transformation 
on $M_4\times Z\ma{N}$. 
Konisi and Saito\cite{KS} pointed out 
that the un-shifted Higgs field  $H_{nk}(x)$
is a connection accompanying the parallel transformation of the fermion
field on the discrete space without recourse of NCG, and therefore
without the introductions of the generalized exterior derivative ${\bf d}$
and the matrix $M_{nk}$ in our formulation. 
Their conclusion is different from ours. That is, in the present
formulation the shifted Higgs field ${\mit \Phi}_{nk}(x)$ is a 
connection on the discrete space. This is apparent from the remembrance
between Eqs.(\ref{2.19}) and (\ref{2.20}) and also the expression of 
the covariant derivative in Eq.(\ref{4.8}). This difference seems 
to come from  differences between two formulations.
\par
The matrix formulation of NCG similar to 
Chamseddine, Felder and Fr\"olich's work\cite{Cham} was also developed
by use of $d$ and $d_\chi$ instead of $\gamma_\mu$ and $\gamma_5$
, by which the shifting rule was confirmed to be reasonable. 
The shifting rule
 is very important in our calculations. The matrix formulation naturally 
 yields the same results as that of the non-matrix approach 
 not only in Yang-Mills-Higgs sector and also in fermionic sector.
\par
Our NCG on $M_4\times Z\ma{N}$ has already applied to reconstruct
$SU(5)$\cite{O1} and $SO(10)$\cite{O10} grand unified theory (GUT)
and the left-right symmetric gauge theory.\cite{LR} Generally speaking, 
the restrictions on the Higgs potential terms and interacting terms
in Eq.(\ref{2.35}) are so stringent that we have to devise the
model constructions of such theory. For $SU(5)$ GUT, the model
construction is fairly successful because the interaction term of 
the adjoint and 5-dimensional Higgs fields yields colored Higgs with GUT 
scale mass. However, the Higgs interacting terms are so limited
that it might contradict with the quantization.
In the case  of $SO(10)$ GUT unusual Higgs field
must be incorporated to maintain the Higgs potential term which 
is responsible for the seesaw mechanism.  
In order to remove such kind of unnaturalness of the reconstruction 
of the various models, we must continue to improve NCG approach.\par
It is still unknown whether the quantization of 
 the Lagrangian derived from NCG in a way 
to be compatible with NCG approach is possible or not.
However, It is confirmed  that the Higgs field is a gauge particle 
with the equal footing to the ordinary gauge field such as weak bosons
in the Weinberg-Salam theory. Thus, it might be expected
that the naturalness problem with respect to the quadratic divergence of 
the Higgs field is removed if one calculates it 
with much attention
on that the Higgs field is a gauge field. If so, the mass relation 
$m\ma{H}=\frac{4}{\sqrt{3}}m\ma{W}\sin\theta\ma{W}$
proposed by the present author\cite{O3} may hold without any correction
in the same way as  $m\ma{W}=m\ma{Z}\cos\theta\ma{W}$.
These circumstances may be investigated by use of the BRST invariant 
Lagrangian of spontaneously broken gauge theory in NCG on the discrete
space $M_4\times Z_2$ recently proposed by the present author\cite{O4} and
Lee, Hwang and Ne'eman.\cite{LHN}

\section*{ Acknowledgements}
The author would like to
express his sincere thanks to
Professors J.~Iizuka,
 H.~Kase, K.~Morita and M.~Tanaka 
for useful suggestion and
invaluable discussions on the non-commutative geometry.
\begin{appendix}
\section{Comparison between NCG in this article
and our previous NCG }
When we reconstructed the standard mode in noncommutative geometry 
on the discrete space $M_4\times Z_2$,\cite{MO1}$^-$\cite{OSM} 
we use only one $\chi$, not
$\chi_1$ and $\chi_2$ appearing 
in this article. Thus, we investigate in this appendix
the relation between two formulations. Let us first expand 
the related expressions in the formulation with $\chi_1$ and $\chi_2$.
From Eq.(\ref{2.13}), the generalized gauge fields are written as
\ba
    \a {\cal A}(x,1)=A_\mu(x,1)dx^\mu+{\mit\Phi}_{12}(x)\chi_2,\nonum
    \a {\cal A}(x,2)=A_\mu(x,2)dx^\nu+{\mit\Phi}_{21(x)}\chi_1.\label{a1}
\ea
From Eq.(\ref{2.25}), the generalized  field strengths are written as
\ba
    \a\hskip-0.5cm {\cal F}(x,1)=\frac12 F_{\mu\nu}(x,1)dx^\mu\wedge dx^\nu
 +D_\mu{\mit\Phi}_{12}(x)dx^\mu\wedge\chi_2+V_{12}(x)\chi_2\wedge\chi_1,\nonum
    \a \hskip-0.5cm{\cal F}(x,2)=\frac12 F_{\mu\nu}(x,2)dx^\mu\wedge dx^\nu
 +D_\mu{\mit\Phi}_{21}(x)dx^\mu\wedge\chi_1+V_{21}(x)\chi_1\wedge\chi_2,
 \label{a2}
\ea
where
\ba
   \a   \hskip-0.8cm  F_{\mu\nu}(x,k) 
     =\p_\mu A(x,k)-\p_\nu A(x,k)+[A_\mu(x,k), A_\nu(x,k)],
     \hskip 0.5cm {\rm for}\quad k=1,2,     \label{a3}\\
\a  \hskip-0.8cm D_\mu{\mit\Phi}_{12}(x)
=\p_\mu{\mit\Phi}_{12}+A_\mu(x,1){\mit\Phi}_{12}(x)
-{\mit\Phi}_{12}(x)A_\mu(x,2),\nonum
\a  \hskip-0.8cm D_\mu{\mit\Phi}_{21}(x)
=\p_\mu{\mit\Phi}_{21}+A_\mu(x,2){\mit\Phi}_{21}(x)
-{\mit\Phi}_{21}(x)A_\mu(x,1), \label{a4}
\ea
and
\ba
   \hskip-1cm    V_{12}(x)
     =&&({\mit\Phi}_{12}(x)+M_{12})({\mit\Phi}_{21}(x)+M_{21})
         -\sum_i a_i^\dagger(x,1)M_{12}M_{21}a_i(x,1),  \nonum
   \hskip-1cm   V_{21}(x)=&&
     ({\mit\Phi}_{21}(x)+M_{21})({\mit\Phi}_{12}(x)+M_{12})
         -\sum_i a_i^\dagger(x,2)M_{21}M_{12}a_i(x,2).  \label{a5}
\ea
In these equation, if $\chi_1$ and $\chi_2$ are equally set to be $\chi$ and
the replacements such as 
$A_\mu(x,1)\to A_\mu(x,+)$, $A_\mu(x,2)\to A_\mu(x,-)$, 
${\mit\Phi}_{12}(x)\to {\mit\Phi}(x,+)$ 
and ${\mit\Phi}_{21}(x)\to {\mit\Phi}(x,-)$ are performed, we can obtain
the equations introduced in the formulation with only one $\chi$
in Refs.\citen{MO1}$\sim$\citen{OSM}.
We can obtain the 
same Lagrangian also in Refs.\citen{MO1}$\sim$\citen{OSM} 
as that in Eq.(2.35).
Therefore, we can say that two formulations are equal.
\end{appendix}
\def\jmp{J.~Math.~Phys.$\,$}
\def\pl{Phys. Lett.$\,$ }
\def\np{Nucl. Phys.$\,$}
\def\ptp{Prog. Theor. Phys.$\,$}
\def\prl{Phys. Rev. Lett.$\,$}
\def\pr{Phys. Rev. D$\,$}
\def\mp{Int. Journ. Mod. Phys.$\,$ }

\end{document}